# A VLSI-Like Design and Verification Flow for Quantum Circuits: A Case Study in PI Estimation

Yu-Ting Kao
Industrial Technology Research Institute
Hsinchu, Taiwan
SunnyKao@itri.org.tw

Yeong-Jar Chang
Industrial Technology Research Institute
Hsinchu, Taiwan
ot@itri.org.tw

***Abstract*—This paper proposes a VLSI-like design and verification flow for quantum circuit implementation, covering four levels: behavioral modeling, register-transfer level (RTL), gate level, and full-circuit integration with quantum algorithms. Using quantum Monte Carlo methods for π estimation as a case study, the proposed flow provides a systematic path from initial specifications to large-scale quantum circuit implementation. The paper highlights two key contributions. First, quantum digital design supports a broad range of applications beyond predefined circuit templates. This capability is particularly important for application-specific operations in algorithms such as Shor's, Grover's, and quantum amplitude estimation (QAE), and can make significant advancement of fault-tolerant quantum computing (FTQC). Second, classical simulation shows that replacing random sampling with grid sampling can achieve quadratic improvement in this π-estimation case without quantum computation. This suggests that the quadratic speedup may arise from the grid sampling strategy rather than from the QAE circuit itself. By adapting conventional VLSI digital design principles, this four-level flow addresses a key bottleneck in quantum circuit design and enables flexible design initiation from multiple entry points, including specifications, RTL, and gate-level descriptions, paving the way for the practical deployment of quantum computing.***

## I. INTRODUCTION

Brassard et al. [1] introduced quantum amplitude estimation (QAE) and established its quadratic query advantage for Monte Carlo estimation. Later studies developed alternative QAE variants and readout methods [2]–[6], along with quantum-arithmetic building blocks [7], [8]. Noto [9] used a quantum squarer to estimate π with QAE, whereas Chang et al. [10] implemented a quantum Monte Carlo circuit that estimates the probability directly. Our benchmark expresses the quarter circle test as synthesizable Boolean logic, allowing verification from behavioral and RTL models through logic synthesis, reversible translation, and QAE integration.

Digital Boolean and arithmetic circuits are fundamental building blocks of Shor factoring [11], Grover search [12], and amplitude estimation [1]. The practical difficulty is to express an application predicate as a reversible, synthesizable circuit to verify its interface with the QAE circuit, which motivates the cross-level verification flow studied here.

Although Monte Carlo estimation of π is not necessarily the most efficient approach for this specific problem, Monte Carlo methods remain widely used and sometimes essential in many applications, particularly when existing software cannot be modified, the problem involves high-dimensional spaces, or other practical constraints limit alternative methods. The π-estimation example is used here primarily to demonstrate the methodology and workflow of quantum digital circuit design, rather than to claim an optimal solution to π estimation itself. The proposed design and verification methodology is general and can be applied to other quantum circuits, including those implementing algorithms such as Shor's, Grover's, and QAE. Although these algorithms are well established, their quantum circuit implementation remains challenging. Application-specific computations, such as modular exponentiation ($11^n$ mod 21), quantum oracles, and Monte Carlo kernels, must be translated from specifications into reversible circuits and carefully verified.

Building on [13], this work extends the existing synthesis and translation flow by addressing technical aspects not fully covered in the original study. It presents a cross-level design and verification methodology for π estimation, covering behavioral modeling, register-transfer level (RTL), gate-level design, and full-circuit integration. This work also highlights two broader implications. First, quantum digital design enables customized computations beyond predefined circuit templates, potentially supporting a broad range of applications. Second, classical simulations comparing random and grid sampling show that quadratic improvement in π estimation can be achieved through grid sampling without quantum circuit execution. These findings clarify both the practical challenges of quantum circuit implementation and the role of sampling strategy in estimation efficiency.

By varying the oracle precision (***b***) and the QAE register size (***q***), we investigate the error and derive the following expression for $\boldsymbol{\pi}$ estimation:

$$\hat{\pi} = 4cos^2\left(\frac{\pi k}{2^q}\right) \quad (1)$$

where ($k$ in $\{0,1,2, \ldots ,2^q-1\}$) denotes the encoded measurement outcome observed most frequently.

The main contribution of this work is a systematic methodology for designing and verifying application-specific quantum circuits, demonstrated through π estimation.

## II. Methodology

Quantum measurement probabilities depend on squared amplitudes, so a common phase applied to an entire state is unobservable in isolation. Boolean logic is embedded in a reversible circuit using ancillas and NOT–CNOT–Toffoli (NCT) gates [14]–[16]. We use b bits per coordinate, $n = 2b$ coordinate qubits, together with one predicate flag, synthesis ancillas, and q counting qubits.

### A. *Effectiveness of QAE*

Direct measurement of Monte Carlo computation does not provide quantum advantages [17], motivating the use of quantum amplitude estimation (QAE). Numerous studies have demonstrated the potential of QAE for a broad range of applications [3]–[6]. Standard QAE estimates a probability to additive error $\varepsilon$ using $O(1/\varepsilon)$ oracle queries, compared with $O(1/\varepsilon^2)$ samples required by classical Monte Carlo methods [1], [2]. With $\boldsymbol{q}$ counting qubits, standard QAE requires up to $2^q-1$ controlled Grover applications [1], [18]. Building on these principles, our experiments adopt a QAE-based approach to achieve a quadratic reduction in sampling complexity.

For circuit implementation, we use a ***q***-bit controlled-Q approach to map the results of Monte Carlo computations onto a unit circle, where the corresponding phase angle is estimated using the quantum Fourier transform (QFT) [19]. The Monte Carlo computation follows a grid-based sampling approach similar to [20], while the quantum convergence method is based on [21], [22].

The main contribution of this work lies in the customized implementation of Monte Carlo computations and a clearer, more accessible quantum circuit design and verification flow. The overall quantum circuit integration does not introduce a new method; instead, established techniques are adopted to validate the practicality of the proposed design methodology.

### B. *VLSI-Like Quantum Design Flow*

The flow adapts VLSI verification practice to quantum circuits. Each stage requires a different comparison as the design moves from Boolean logic to a reversible circuit and then into QAE. Fig. 1 shows the nine stages.

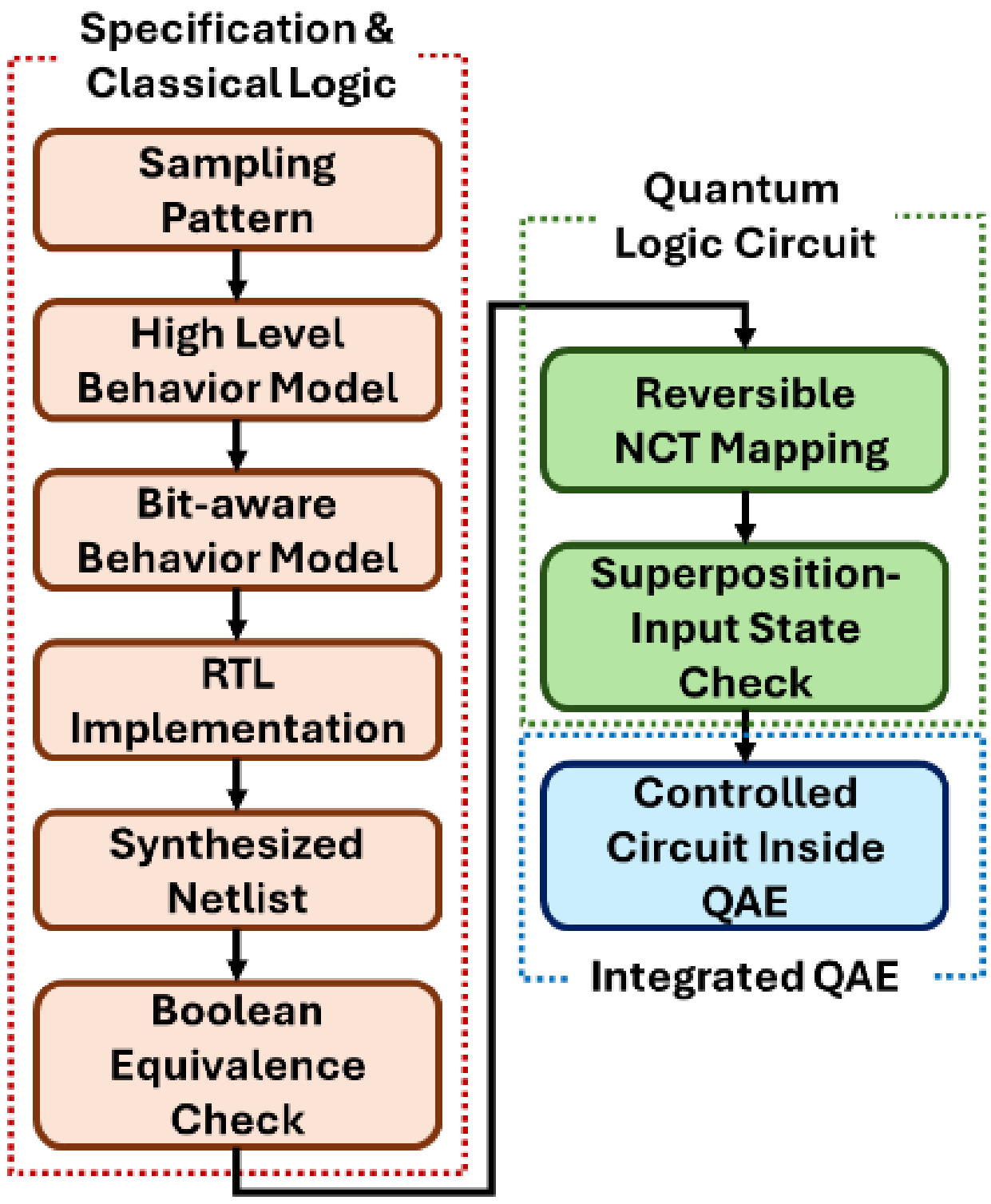

Fig. 1. Nine-stage verification flow grouped into four design levels: behavioral (Stages 1–3), RTL (Stage 4), gate (Stages 5–8, VLSI and quantum gates), and full circuit (Stage 9).

The nine stages are grouped into the four design levels of a conventional VLSI flow, as listed in Table I.

(i) Behavioral level (Stages 1–3): the sampling pattern and the high-level and bit-aware behavioral models describe the computation with vectors and matrices instead of individual gates, so the design can be simulated and debugged before any gate is implemented.

(ii) RTL level (Stage 4): the bit-aware model is written in Verilog (Listing 2 shows $b = 2$).

(iii) Gate level (Stages 5–8): this level covers both VLSI gates and quantum gates. Stages 5–6 produce the synthesized standard-cell netlist and check its Boolean equivalence, as in a conventional VLSI flow. Stages 7–8 translate each VLSI gate into NCT quantum gates one by one and verify the reversible circuit on basis states and on a designated superposition input.

TABLE I DESIGN LEVEL, VERIFICATION TARGET, AND PASS CRITERION AT EACH STAGE

| **Design level** | **Stage** | **Objects compared** | **Check / pass criterion** |
|---|---|---|---|
| Behavioral | 1 | Sampling pattern | Grid resolution b and midpoint rule fixed; defines the encoded probability $p_b$ |
| | 2–3 | High-level and bit-aware behavioral models | Same predicate value for every encoded input |
| RTL | 4 | RTL implement ation | Verilog of the bit-aware model; Boolean function compared at Stage 6 |
| Gate (VLSI) | 5 | Synthesize d netlist | Standard-cell netlist generated by logic synthesis; Boolean function compared at Stage 6 |
| | 6 | Netlist vs. bit-aware model | Identical Boolean function over all $2^n$ inputs |
| Gate (quantum) | 7 | Reversible (NCT) mapping on basis states | Correct output bit; ancillas initialized and returned clean, no residual garbage |
| | 8 | Designated superpositi on input | Output states agree up to one common global phase (that input only) |
| Full circuit | 9 | Controlled circuit inside QAE | Counting-register distribution matches the readout convention of the implemented Grover operator |

(iv) Full-circuit level (Stage 9): the verified gate-level circuit is integrated with the quantum algorithm (here, the state-preparation operator A, the controlled Grover powers, and the inverse QFT of QAE) and checked as a whole.

As in VLSI design, a behavioral model can represent a block before its gate-level implementation is available, and be replaced once the implementation is verified. Similarly, mathematical and logical operations implemented in Python can serve as behavioral models to simplify quantum circuit verification.

A gate-level simulator processes each gate sequentially, requiring a full-state pass per gate. In contrast, a behavioral simulator executes the block in one step. For the π-estimation block, it flips flag F if grid point (x, y) lies inside the quarter circle. Thus, simulation cost no longer scales with gate count.

In a controlled operation, a global phase becomes a relative phase between the active and inactive branches, so circuits that pass the same standalone check may produce different QAE results; Stage 9 checks this behavior in the integrated circuit. The implementation follows [13], from behavioral modeling through Verilog and Design Compiler to gate-by-gate NCT translation [14]–[16], [23]–[26].

The proposed flow is generally applicable to FTQC implementations of Shor's, Grover's, Brassard's, and other quantum algorithms. The following section demonstrates the flow using Monte Carlo-based π estimation.

### C. *π Estimation Circuit Design*

In our case study, the Monte Carlo method samples points within a two-dimensional square. As the sample set becomes sufficiently large and dense, the probability of a point falling inside the quarter circle approaches the ratio of the circle's area to the square's area:

$$p = \frac{\left[\left(\frac{1}{4}\right)\pi r^2\right]}{r^2} = \frac{\pi}{4} \Rightarrow \pi = 4p$$

By estimating this probability using quantum computing and the QAE algorithm, we can estimate the value of π. Due to the computational cost of exploring large sampling spaces, classical computing typically relies on random sampling, as illustrated in Fig. 2(a). In contrast, quantum computing can exploit superposition to encode multiple sampling points in a single quantum state, enabling grid-based sampling, as illustrated in Fig. 2(b). The proposed VLSI-like design flow is used to implement the quantum judgement circuit that determines whether each sampling point lies inside the quarter circle.

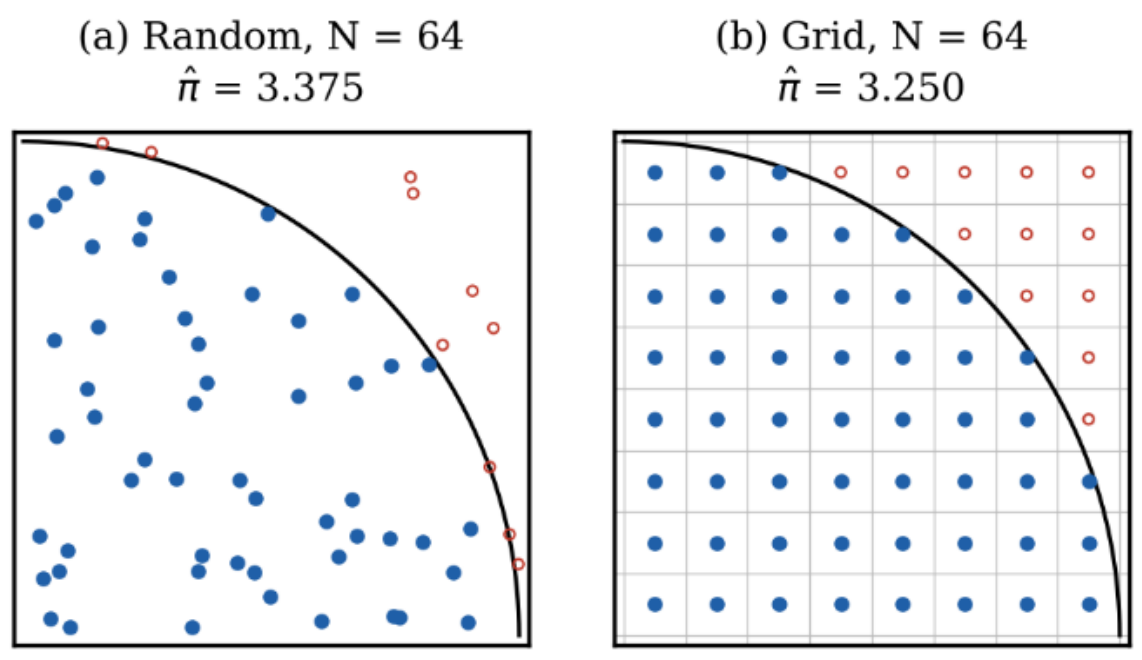

(a) (b)

Fig. 2. Random (a) and grid (b) sampling with the same N = 64 points; filled markers lie inside the quarter circle.

```
import numpy as np

# (1) Monte Carlo + QAE: judgement circuit of Listing 2
b = 2; r = 2**b
x, y = np.divmod(np.arange(r * r), r)
F = np.where((x + 0.5)**2 + (y + 0.5)**2 <= r**2, 1, 0)
pi_hat = 4 * np.mean(F)

# (2) Shor: factor 21 from the period of f(x) = 11^x mod 21
np.period = lambda f: np.nonzero(f[1:] == f[0])[0][0] + 1
f = 11**np.arange(12) % 21
T = np.period(f)
factors = np.gcd(11**(T // 2) + [-1, 1], 21)

# (3) Grover: search for marked items / minimum finding
marked = np.nonzero(F == 0)[0]
f_min = np.minimum.reduce(f)
```

Listing 1. Behavioral Models of three Quantum Circuits

Listing 1. consists of three blocks. Block (1) evaluates the judgement condition for all grid points at once with np.where, and np.mean gives the probability estimated by QAE. Block (2) represents Shor's algorithm: the custom np.period finds the period of 11^x mod 21, from which np.gcd yields the factors 7 and 3. Block (3) represents Grover's algorithm: np.nonzero searches for the marked (outside) points, and np.minimum performs minimum finding. Each quantum algorithm is thus expressed in one or two vectorized NumPy statements, allowing its function to be verified before any gate is implemented.

First, the behavior model in Listing 1 replaces complex quantum gate operations with a simple Python condition, $(x + 0.5)^2 + (y + 0.5)^2 < r^2$, enabling early functional verification of Monte Carlo methods. A key distinction between quantum behavior models and conventional models lies in leveraging NumPy's vectorization capabilities, enabling array-based operations to replace explicit Python for loops. This approach provides an efficient means of representing repetitive computations in quantum computing. NumPy functions include np.where, np.select, and np.nonzero, among others.

More notably, we define np.period as a custom function, together with np.minimum and np.mean, to provide conceptual abstractions for the computational capabilities of Shor's algorithm, Grover's algorithm, and Brassard's QAE, respectively. These functions serve as behavioral interfaces for representing the corresponding quantum algorithms within the proposed model.

By comparing random and grid sampling, we demonstrate a quadratic speedup in classical simulation, showing that the speedup originates from the sampling strategy rather than the QAE circuit itself.

Once the functionality is verified, the next step is quantum circuit implementation. However, the quantum state space grows exponentially with the number of qubits. For example, a six-qubit circuit represents (N=64) computational basis states, making circuit implementation challenging. To address this challenge, we propose a VLSI-like RTL design flow in which Design Compiler synthesizes the logic implementing the Python condition in the behavior model, which is then translated into NCT gates. This approach enables systematic circuit generation without explicitly enumerating the exponentially growing state space during logic synthesis. To implement the condition

$$(x + 0.5)^2 + (y + 0.5)^2 < r^2$$

we multiply both sides by four to obtain

$$(2x + 1)^2 + (2y + 1)^2 < (2r)^2$$

```
module calculate_PI_V2_2bit(
    input [1:0] X,
    input [1:0] Y,
    output F
);
wire [2:0] XS, YS;
wire [6:0] XS_squared, YS_squared, dis;

assign XS[2:1] = X[1:0];
assign XS[0] = 1'b1;
assign YS[2:1] = Y[1:0];
assign YS[0] = 1'b1;
assign XS_squared = XS * XS;
assign YS_squared = YS * YS;
assign dis = XS_squared + YS_squared;
assign F = ~dis[6];

endmodule
```

Listing 2. RTL code of judgement circuit

In Listing 2, each coordinate is left-shifted by one bit, and the least significant bit is set to 1, yielding $(XS = 2x + 1)$ and $(YS = 2y + 1)$.

Let ***dis*** denote the sum of their squares. If ***dis***<(2r)$^2$, the comparison result indicates that the sampling point lies inside the quarter circle. For the specified bit width and threshold, this condition is encoded by the absence of a carry into the most significant bit. Therefore, the flag is defined as F=~***dis***[6], indicating whether the sampling point lies inside the quarter circle.

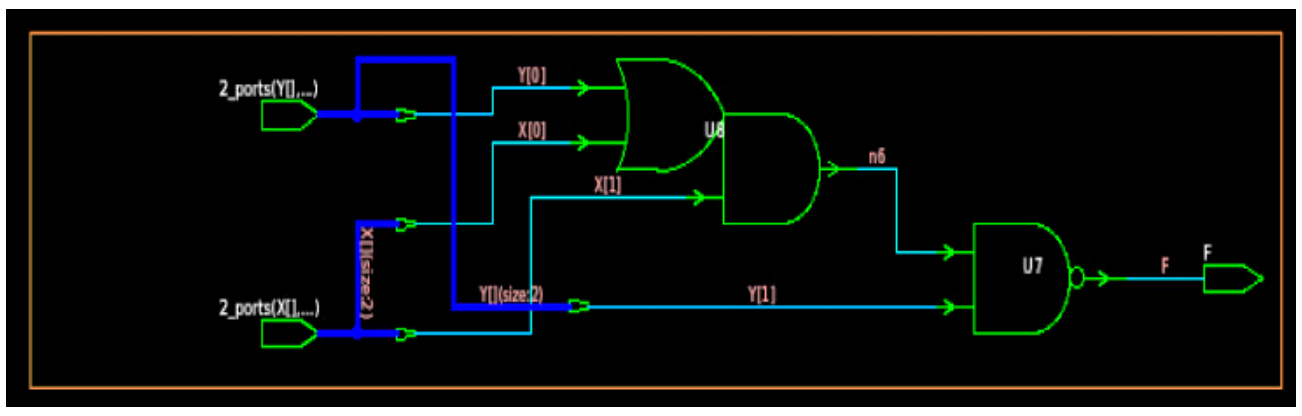

Fig. 3. Gate-level netlist synthesized from Listing 2 by Design Compiler.

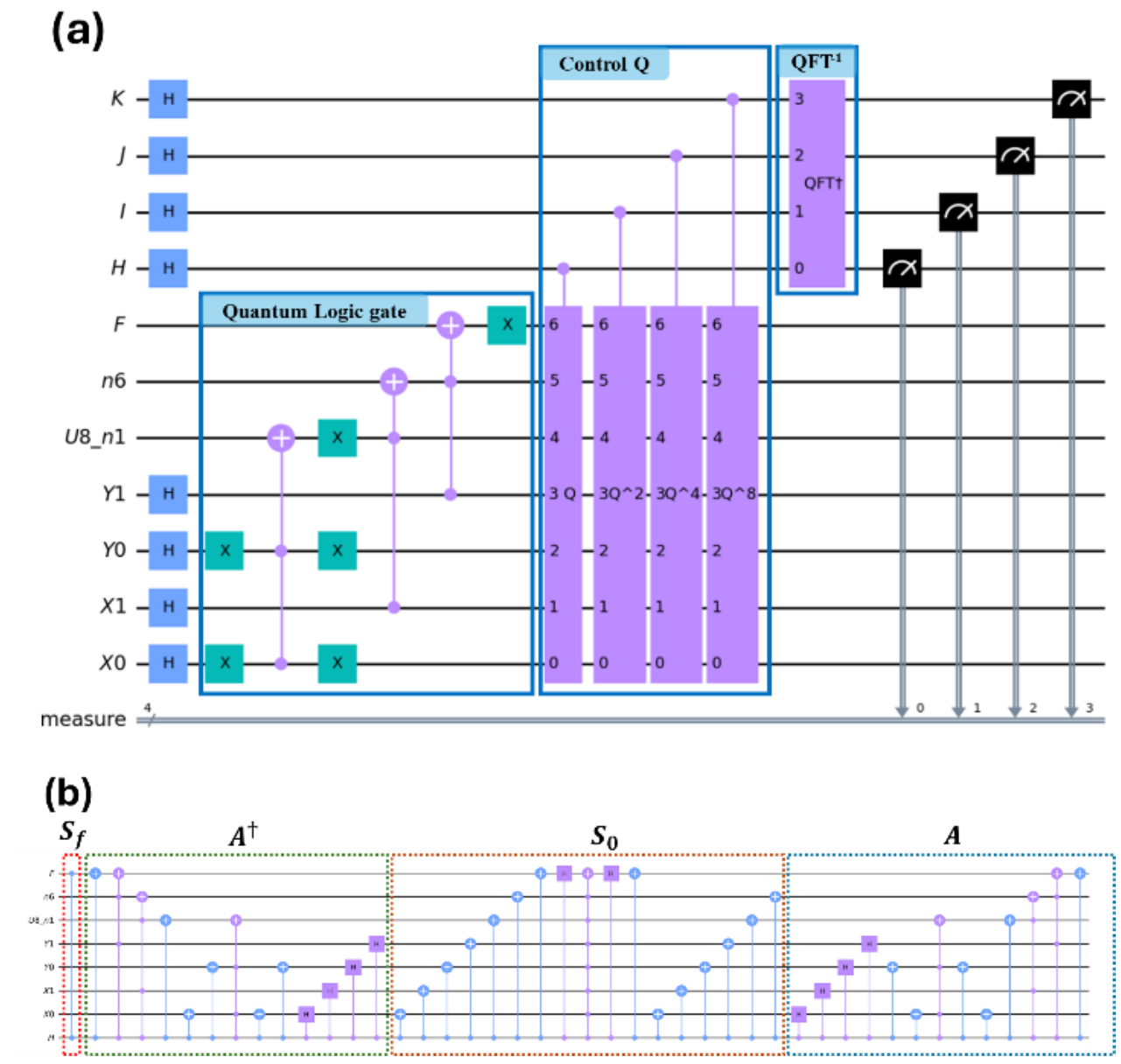

Fig. 4. (a) The overall QAE circuit. (b) A single controlled-$Q_{cir}$ operation.

For subsequent synthesis and verification, we set the total number of grid points to (N=16). The RTL is first simulated with a testbench that applies all 16 inputs; 13 of them give $F$ = 1, so $\pi_b$ = 4 × 13/16 = 3.25, in agreement with the behavioral model. The RTL is then synthesized with Design Compiler into one OR, one AND, and one NAND gate (Fig. 3) and checked for equivalence, using standard VLSI tools and file formats. Using ITRI's translator, the VLSI gate-level implementation is mapped to quantum NCT gates. Together with the ancilla signals (U8_n1) and (n6), these gates implement the judgement circuit, denoted as the ***A*** operator in Fig. 4(a). For π estimation, an output state of |1⟩ indicates that a grid point lies inside the quarter circle. The QAE algorithm estimates the corresponding probability (p), from which π is obtained as π = 4P. Therefore, the ***A*** operator must be integrated into the QAE circuit to perform π estimation.

In QAE circuit implementations, the following operator definitions are commonly adopted:

$$S_f = I - 2\Pi_{\text{good}}, \qquad S_0 = I - 2|0\rangle\langle 0|. \quad (2)$$

$$Q_{\text{ref}} = -AS_0A^\dagger S_f, \quad Q_{\text{cir}} = AS_0A^\dagger S_f = -Q_{\text{ref}}. \quad (3)$$

Here, $\boldsymbol{Q_{ref}}$ denotes the Grover operator as defined by the reference formulation, whereas $\boldsymbol{Q_{cir}}$ denotes the operator implemented by the quantum circuit.

This $\boldsymbol{S_0}$ equation denotes a phase flip of the amplitude associated with the all-zero basis state, $|0^n\rangle$, while leaving all other basis states unchanged. To implement this operator, apply X gates to map the all-zero state to the all-one state, apply a multi-controlled Z gate to flip its phase, and then restore the original basis states using X gates, ensuring that only the all-zero state undergoes a phase flip.

The $\boldsymbol{S_f}$ equation denotes a phase flip of the amplitudes associated with the target states, while leaving all other basis states unchanged. To implement this operator, use a reversible oracle to identify the target states via a flag qubit, then apply a flag-controlled Z gate to flip only their phases.

Fig. 4 shows the resulting QAE circuit for b = 2 and q = 4. In Fig. 4(a), $A$ acts as the judgement circuit to generate the flag $F$; the counting qubits (K, J, I, H) control the Grover powers; and QFT† converts the phase into the counting-register outcome. Fig. 4(b) expands one controlled $\boldsymbol{Q_{cir}}$ into $\boldsymbol{S_f}$, $\boldsymbol{A}^\dagger$, $\boldsymbol{S_0}$, and ***A***, corresponding to (3).

## D. *Interpreting the QAE Readout*

The QAE readout follows three relations under the synthesized Grover convention:

$$p = \sin^2(\theta) \quad (4)$$

$$2\pi\varphi = \pi - 2\theta \quad (5)$$

$$k = 2^q\varphi \quad (6)$$

Eq. (4) follows from the state prepared by A. Writing $A|0\rangle = \cos\theta\,|\psi_{\text{bad}}\rangle + \sin\theta\,|\psi_{\text{good}}\rangle$, the probability of measuring the flag F = 1 is $p = \sin^2\theta$, so estimating p is equivalent to estimating the angle θ.

The operator $\boldsymbol{S_f}$ reflects the state about the unmarked-state axis $|\psi_{\text{bad}}\rangle$, whereas $\boldsymbol{AS_0A^\dagger}$ reflects it about the axis orthogonal to the prepared state $|\psi\rangle = \boldsymbol{A}|0\rangle$. The composition of these two reflections yields a rotation by $2\boldsymbol{\theta}$ with an overall negative sign. Consequently, the operator implemented by the quantum circuit, $\boldsymbol{Q_{cir}}$, differs from the reference operator, $\boldsymbol{Q_{ref}}$, by a global negative sign, such that $Q_{cir} = -Q_{ref}$. Since the eigenvalues of $Q_{ref}$ are $e^{\pm i2\theta}$, those of $Q_{cir}$ become $-e^{\pm i2\theta} = e^{i(\pi\pm 2\theta)}$. Thus, the additional negative sign introduces a phase shift of π. Let φ denote the normalized phase estimated by the phase-estimation register, such that an eigenvalue is expressed as $e^{i2\pi\varphi}$. Equating the corresponding phase yields $2\pi\varphi = \pi - 2\theta$, as shown in Eq. (5). This derivation establishes the relationship between the global negative sign in the synthesized operator and the phase estimated by the QAE circuit.

Eq. (6) follows from how the counting register reads φ. The Ctrl-Q circuit applies successive controlled powers of Q, namely $Q^0, Q^1, Q^2, \ldots, Q^{15}$ to generate phase factors corresponding to 0φ, 1φ, 2φ, 3φ, ..., 15φ across the corresponding computational-basis components of the phase-estimation qubits. The inverse quantum Fourier transform (QFT†) then maps the accumulated phase information to computational-basis states, allowing the phase to be estimated by measuring the register. The most frequently observed measurement outcome, k ∈ {0, 1, 2, ..., 15}, provides an estimate of the phase according to $\varphi \approx k/2^q$, or equivalently, $k \approx 2^q\varphi$. In this design, q=4. When φ is exactly representable on the q-bit phase grid, the equality $k = 2^q\varphi$ holds exactly; otherwise, k is the nearest phase-grid estimate, subject to the measurement probability distribution.

Combining (4)–(6) yields the general π estimator

$$\hat{\pi} = 4\hat{p} = 4sin^2(\theta) = 4sin^2[(\pi - 2\pi\varphi)/2]$$
$$= 4\cos^2(\pi\varphi) = 4\cos^2(\pi k/2^q) \quad (7)$$

For example, when b=2, the encoded probability is p=13/16, yielding θ ≈ 1.123 rad, φ ≈ 0.143, and k ≈ 2.28 for q = 4. However, QAE only returns integer-valued outcomes. The nearest outcome, k=2, gives $\hat{p} = \cos^2(\pi/8) \approx 0.854$ and $\hat{\pi} \approx 3.4142$.

Monte Carlo analysis typically extracts meaningful information through statistical aggregation over numerous samples. In this work, quantum-based π estimation serves as a case study to demonstrate design principles across multiple abstraction levels. The proposed VLSI-like design flow is generalizable to other applications.

However, deriving π directly from the QAE measurement outcome ***k*** is problematic when the computation relies on a prior approximation of π (e.g., 3.1416) and corresponding trigonometric evaluations. This introduces a circular dependency, using the assumed value to estimate the same unknown quantity.

This paper resolves this issue by deriving a complete mathematical formulation, Eq. (7), that combines trigonometric identities to estimate π without relying on a predefined approximation, thereby eliminating the circular dependency and establishing a mathematically consistent framework for quantum-based π estimation.

## III. Results

Since the VLSI-like design methodology enables efficient implementation of the π-estimation circuit, the next step is to evaluate the accuracy of π estimation using the quantum circuit. The parameter ***b*** denotes the number of bits used to represent the (x,y) coordinates, whereas ***q*** denotes the number of QFT estimation qubits. Both parameters affect the estimation accuracy and convergence behavior differently.

Discretization introduces deterministic bias rather than sampling variance, so repeated runs cannot average it away, and the bias is non-monotone in the grid size (Table II). Because the state-preparation circuit ***A*** encodes the grid, any effect of the sampling pattern occurs before QAE, and increasing ***q*** therefore cannot reduce this bias floor.

TABLE II DETERMINISTIC MIDPOINT-GRID APPROXIMATION VS. RANDOM SAMPLING

| b (points) | *Estimate* | \|Estimate − π\| | *Random-sampling std. error* |
|---|---|---|---|
| **2 (16)** | **3.25** | **0.10841** | **0.41055** |
| **4 (256)** | **3.1719** | **0.03028** | **0.10264** |
| **6 (4,096)** | **3.1475** | **0.00587** | **0.02566** |
| **8 (65,536)** | **3.1417** | **0.00007** | **0.00641** |
| **10 ($2^{20}$)** | **3.1417** | **0.00009** | **0.00160** |
| **12 ($2^{24}$)** | **3.1416** | **0.000004** | **0.00040** |

Table II reports the deterministic midpoint-grid approximation for b = 2 to 12. The reported error is $|\pi_b - \pi|$. Odd values of b are omitted from Table II for space; for example, b = 2 and b = 3 give the same estimate 3.25, so the error is non-monotone across resolutions, as expected from lattice-point counting [20]. Over b = 2 to 8, a log–log fit of the peakgrid error gives a slope of −0.93, close to $O(1/N)$, whereas the random-sampling standard error in the last column decreases as $O(1/\sqrt{N})$; at b = 8 the grid error is about 90 times smaller than the random-sampling standard error.

Fig. 2 makes this comparison concrete. In Fig. 2(a), 54 of 64 random points fall inside the quarter circle, giving $\hat{\pi} = 4 \times 54/64 = 3.375$ for this particular draw; another draw gives a different value, with a standard deviation of 0.21 at $N = 64$. In Fig. 2(b), 52 of the 64 grid points ($b = 3$) fall inside, giving $\pi_b = 3.25$ on every run. Random points cluster in some regions and leave gaps in others, so the estimate depends on chance; grid points cover the square evenly, and the only error comes from the cells cut by the arc, whose over- and under-counts partly cancel. Extending the fit to $b$ = 12 gives a slope of −0.85 for the grid error, against −0.50 for random sampling: when the number of points grows 100-fold, the random-sampling error shrinks only 10-fold, whereas the grid error shrinks about 50-fold. In this two-dimensional example, replacing random points with a regular grid therefore already provides the quadratic improvement usually attributed to QAE, without any quantum operation.

Let $P(k)$ denote the probability of observing outcome $k$, and let $M = 2^q$ .The folded probability combines each pair of symmetric raw outcomes, counting each of the two self-conjugate endpoints only once:

$$P_{\text{pair}}(k) = \begin{cases} P(k) + P(M-k), & 0 < k < M/2, \\ P(k), & k = 0 \text{ or } k = M/2. \end{cases} \quad (8)$$

The simulation results are presented in Table III. Errors measured relative to π include the grid discretization bias. For $q$=4, the mathematical readout convention of $Q_{ref}$ yields dominant measurement bins {6,10}, whereas the circuit-based readout convention of $Q_{cir}$ yields {2,14}. This eight-bin shift can only be identified by accounting for the global phase in the comparison, highlighting the importance of full-circuit verification (Stage 9).

In Table III, $k$ denotes the measurement bin ($0 \leq k \leq M/2$) with the largest $P_{pair}(k)$.

TABLE III COUNTING-REGISTER SWEEP AT $b$ = 2 FROM THE EXACT DISTRIBUTION

| $q$ | $k$ | $\hat{\pi}$ | error vs $\pi_b$ | *error vs π* | $P_{pair}(k)$ |
|---|---|---|---|---|---|
| **4** | **2** | **3.4142** | **5.053%** | **8.68%** | **0.771** |
| **6** | **9** | **3.2688** | **0.578%** | **4.05%** | **0.951** |
| **8** | **36** | **3.2688** | **0.578%** | **4.05%** | **0.417** |
| **9** | **73** | **3.2497** | **0.009%** | **3.44%** | **0.999** |

At fixed b = 2, the probability of a point falling inside quarter-circuit is p=13/16, yielding $\pi_b$ = 4p = 3.25. The peak-bin error relative to the encoded estimate is 5.053% at q = 4, approximately 0.578% at q = 6–8, and 0.009% at q = 9. Its non-monotonicity reflects phase-bin alignment. As $q$ increases, the estimated π approaches $\pi_b$, with the error relative to the true value of π approaching the 3.45% grid-bias floor. Thus, increasing $q$ alone cannot improve the estimate beyond the accuracy of $\pi_b$.

The results further demonstrate the effect of grid resolution. At b=2 and q=9, QAE yields $\hat{\pi}$ =3.2497, within 0.009% of the encoded grid estimate. Increasing the grid resolution to b=8 and using q=10 gives $2^q\varphi \approx 156.99$, as calculated from Eq. (4)–(7). Because this value lies almost exactly on an integer estimation bin, the predicted peak-bin readout is $\hat{\pi}$ =3.14156, within 0.001% of π.

Finally, behavioral and gate-level simulations produce identical output probabilities for all circuits evaluated at both levels, consistent with the mathematical predictions. The behavioral model also supports simulation of the complete 27-qubit circuit for b=8 and q=10, comprising 16 coordinate qubits, one flag qubit, and ten counting qubits. These results support a design workflow in which circuits are debugged at the behavioral level before individual blocks are replaced with verified gate-level implementations. Nevertheless, the gate-level verification steps in Stages 7–9 remain essential before deployment.

## IV. CONCLUSION

This work demonstrates the feasibility of applying conventional VLSI design principles to quantum circuit development through a π estimation case study. By implementing and verifying QAE circuits at the behavioral, RTL, gate, and full-circuit levels for counting precisions q = 4 to 9, we confirm that the proposed methodology can support the systematic development of increasingly complex quantum computations.

This paper presents the principles and implementation details of QAE in an accessible manner, making the underlying technology easier to understand. Digital quantum circuits are synthesized and verified using a VLSI-like design flow, whereas non-digital quantum circuits are constructed using algorithm-specific templates, enabling efficient end-to-end quantum circuit implementation. This paper also derives a mathematical formulation to resolve the deadlock problem arising from using an assumed value of π to estimate π itself. QAE converges to the encoded grid estimate ($\hat{\pi}$ = 3.2497 at b = 2 and q = 9), and with a finer grid the behavioral-level simulation reaches $\hat{\pi}$ = 3.14156 at b = 8 and q = 10. The simulations therefore highlight the importance of distinguishing the effects of the sampling strategy from those of quantum computation.

In the upcoming FTQC era, the proposed methodology can accelerate quantum circuit development by streamlining the translation of computational specifications into verifiable quantum circuits, enabling rapid prototyping for applications such as AI, smart factory, financial optimization, and biotechnology.